\documentclass{iopart}
\usepackage{graphicx}
\usepackage{subfigure}

\newcommand{\ds}{\displaystyle}
\newcommand{\scs}{\scriptscriptstyle}
\newcommand{\pr}{\scriptscriptstyle \prime}

\def\d{\delta}
\def\g{\gamma}
\def\G{\Gamma}
\def\a{\alpha}
\def\o{\omega}

\def\s{\sigma}

\def\DLE{{\cal D}}

\begin{document}

\title[Slow energy relaxation and localization in 1D lattices]
{Slow energy relaxation and localization in 1D lattices}
\author{F Piazza\dag\ddag\footnote[3]
                                {On leave from:
Heriot--Watt University, Physics Department, Edinburgh EH14 4AS, U.K.}, 
        S Lepri\P\ddag, R Livi\dag\ddag} 
\address{\dag Dipartimento di Fisica, L.go E. Fermi 5,
         50125 Firenze, Italy}
\address{\ddag Istituto Nazionale di Fisica della Materia (INFM), 
         Unit\`a di Firenze}
\address{\P Dipartimento di Energetica ``S. Stecco'', Via S. Marta 3, 
         50139 Firenze, Italy}

%\ead{piazza@phy.hw.ac.uk}

\begin{abstract}

We investigate the energy relaxation process produced by thermal baths at zero
temperature acting on the boundary atoms of chains of classical anharmonic
oscillators.  Time--dependent perturbation theory allows us to obtain an
explicit solution of the harmonic problem: even in such a simple system
nontrivial features emerge from the interplay of the different decay
rates of Fourier modes.  In particular, a crossover from an exponential to an
inverse--square--root law occurs on a time scale proportional to the system size
$N$. A further crossover back to an exponential law is observed only at much
longer times (of the order $N^3$). In the nonlinear chain,  the relaxation
process is initially equivalent to the harmonic case over a wide time span, as
illustrated by simulations  of the $\beta$ Fermi--Pasta--Ulam model. The
distinctive feature is that the second crossover is not observed due to the
spontaneous appearance of breathers, i.e. space--localized time--periodic
solutions, that keep a finite residual energy in the lattice. We discuss the
mechanism yielding such solutions and also explain why it crucially depends on
the boundary conditions.

\end{abstract}

\bigskip
{\small submitted to: J. Phys. A, Math. Gen.}

\smallskip
{\small PACS: 63.20.Ry, 63.20.Pw}
%\submitto{\JPA}
%\pacs{63.20.Ry, 63.20.Pw}

\bigskip
{Corresponding author: Francesco Piazza, \texttt{piazza@phy.hw.ac.uk}}
\maketitle

\section{Introduction}

The concept of discrete breathers (DB) has been brought to the foreground
by recent studies on the dynamical properties of anharmonic
lattices~\cite{breath1}. DB are space--localized time--periodic
solutions, with frequencies lying out of the linear spectrum and they
have been proved to exist in a wide class of models~\cite{breath2}. 
Although a fairly large amount of rigorous and numerical work has been 
devoted to the identification of the conditions for their existence
and stability, much less is known about their possible relevance for the 
thermodynamic properties of lattices, both at and out of equilibrium. 
The latter case includes transient dynamics, e.g. relaxation to 
equipartition ~\cite{Cretegny}, and stationary heat transport~\cite{heat}.

An important example of the role played by DB in determining 
relaxation properties has been provided by Tsironis and Aubry~\cite{Aubry1}.
They have shown that a characteristic slow relaxation behaviour of the
energy emerges when a nonlinear chain prepared in a typical
thermalized state, corresponding to a given finite temperature, is put in
contact with a cold bath. This physical setup can be simulated by a damping 
term acting on a number of boundary particles. 
The system eventually reaches a state where a residual finite amount
of the initial energy is kept under the form of DB.  In a subsequent numerical 
work~\cite{Aubry2} it has been further argued that the energy relaxation 
should obey a stretched--exponential law in time. 

Although such a behaviour is considered on its own a signature
of complex dynamics, its origin remains not 
completely understood, even in the more widely studied problem of glassy
relaxation~\cite{Bunde}. Loosely speaking, it is usually
attributed to a multiple--local--minima structure in the energy landscape
of the system. This would originate metastable states capable of trapping
the configuration of the system during the relaxation, thus giving rise to
a very slow decay. In some cases it is also assumed that this behaviour is 
the result of the competition  between two (or among many) pure exponential
processes~\cite{Bunde}. Otherwise, it has been proposed to stem from a 
complex activation mechanism, whereby different degrees of freedom
``defreeze'' at later and later stages. This constitutes a hierarchical
scheme, with faster degrees of freedom successively constraining the slower
ones~\cite{Palmer}.

Beyond the possible relevance of these different scenarios, it
is important to have a definite quantitative description of the
relaxation phenomenon in connection with energy localization.
With this paper we aim to settle these issues on a firm
basis, showing also that the picture is more subtle than one might 
naively expect. As a first step, we consider the simple but
very instructive case of the harmonic chain, with damping acting on its
edge particles. This exercise shows indeed how a slower--than--exponential
relaxation arises as a global effect from the existence of 
different time scales of the Fourier modes. As a matter 
of fact, only a linear system with damping on each particle would display 
simple exponential relaxation. More specifically, a perturbative calculation 
reveals that the energy decay law exhibits a 
first crossover from an exponential to an inverse--square--root law. 
The corresponding crossover time $\tau_0$ is inversely proportional to 
the product of the damping constant times the fraction of damped particles. 

These results turn out to be extremely useful to be compared with the
nonlinear case. Remarkably, the only difference concerns
the long--time behaviour of the decay process. Indeed, in the linear system we 
observe a second crossover back to an exponential law, while in the 
nonlinear one the energy converges to a finite asymptotic value, due to 
spontaneous localization of the energy in the form of DB. However, we find 
that finite--size effects both in space and time may
prevent energy localization, making the interpretation of the energy decay
process more subtle and interesting.

The paper is organized as follows. In section~\ref{meth_mod} we describe
the general features of the relaxation dynamics 
and illustrate the typical layout of numerical simulations.
The results obtained for the linear system are discussed in 
section~\ref{sec_harm}; 
section~\ref{sec_nln} is devoted to the study of the relaxation process
in the $\beta$ Fermi--Pasta--Ulam (henceforth FPU) model ({\em e.g.}
a quartic nonlinearity in the interparticle potential) and is known to 
allow for the existence of breather states.  Finally, we summarize 
and draw our conclusions in section~\ref{conclu}.

\section{\label{meth_mod} Generalities of the relaxation dynamics}

We consider a homogeneous chain of $N$ atoms of mass $m$ and denote with
$u_p$ the displacement of the $p$-th particle from its equilibrium 
position at time $t$ (for the sake of simplicity in what follows we omit 
the explicit dependence on time). The atoms are labelled by the integer 
space index $p = 0, 1, \dots N-1$.
In the nearest--neighbour approximation of the
interactions the equations of motion read
\begin{equation}
\label{eq_mot}
\fl \ \ \ \ \ 
m \ddot{u}_p = V'(u_{p+1}-u_p) - V'(u_p-u_{p-1})  
               - m \sum_{p^{\pr}=0}^{N-1} \G_{p p^{\pr}} 
                        \dot{u}_{p^{\pr}},
\end{equation}
where $V(x)$ is the interparticle potential, for which we assume that
$V'(0)=0$, denoting with $V'(x)$ the derivative of 
$V(x)$ with respect to $x$. The last term in
eq.~(\ref{eq_mot}) represents the interaction of the atoms 
with a ``zero temperature'' heat
bath in the form of a linear damping, characterized by the coupling 
matrix $\G$. 
In what follows we consider the case in which dissipation acts only on the
atoms located at the chain edges, so that $\G$ has the form
\begin{equation}
\label{Gamma}
\Gamma_{p p^{\pr}} = \g \d_{p^{\pr},p} \left[ \d_{p^{\pr},0} +
                                       \d_{p^{\pr},N-1} \right]\quad,
\end{equation}
where $\g$ is the damping constant and $\d_{p^{\pr},p}$ is the
Kronecker delta. As we are dealing with systems in a finite volume we 
impose either free--end ($u_{-1}=u_{0}$, $u_{N-1}=u_{N}$) or fixed--end 
($u_{-1} = u_{N} = 0$) boundary conditions (BC).

The numerical investigations reported in this paper have been performed
according to the following procedure. We integrate
equations~(\ref{eq_mot}) starting from a microcanonical equilibrium condition 
corresponding to a given energy density $E(0)/N$. This condition is obtained
by letting the hamiltonian system ($\g = 0$) perform a short 
microcanonical transient, whose duration is typically ${\cal O} (10^3)$ 
time units. The initial condition for such transient is
assigned by drawing momenta from a Gaussian distribution corresponding to the
desired energy, while all displacement variables $u_p$  are set to zero. 
During the hamiltonian  transient the equations of motion are integrated  
by means of a 5--th order symplectic Runge--Kutta--Nystr\"{o}m  
algorithm~\cite{Calvo}, while a standard 4--th order Runge--Kutta algorithm 
is used for the dissipative  dynamics.  

It is useful to introduce the so--called  symmetrized site energies
\[
e_p \; = \;\frac{1}{2} m \dot{u}_p^2 + 
         \frac{1}{2} \left[ V(u_{p+1}-u_p) + V(u_p-u_{p-1}) \right].
\]
The istantaneous total energy of the system is then given by
\begin{equation}
E = \sum_{p=0}^{N-1} e_p.
\label{energy}
\end{equation}
Due to the presence of dissipation, the initial energy $E(0)$  
decreases in time. Since we are interested in determining the 
decay law of the normalized quantity $E(t)/E(0)$, it is convenient to 
introduce the following indicator  
\begin{equation}
\DLE (t) = \log [ - \log (E(t)/E(0)) ].
\end{equation}
By plotting $\DLE$ versus time in a log--lin scale, a
streched--exponential law of the form  $E(t)/E(0) = \exp[-(t/\tau)^\s]$
becomes a straight line with slope $\beta$ that intercepts the $y$--axis 
at $-\beta\log (\tau)$. As we shall see later on, this representation is very
useful for identifying pure exponential regimes ($\s=1$) and the crossover 
to slower decay laws.

Since we are also interested in the effects of DB formation on the 
relaxation process, we introduce a localization parameter 
${\cal L}$, which provides a rough estimate of the 
degree of energy localization in the system. We define it as  
\begin{equation}
{\cal L}(t) = N \, \frac{\ds \sum_p e_p^2(t)}
                     {\ds \left[ \sum_p e_p(t) \right]^2}.
\label{local}
\end{equation}
Accordingly, the fewer sites the energy is
localized onto, the closer ${\cal L}$ is to $N$. On the other hand, the
more evenly the energy is spread on all the particles, the closer
${\cal L}$ is to a constant of order 1. For instance, in the latter
case ${\cal L}$ for the FPU potential lies within the interval 
[7/4, 19/9], the two extremes being the energy--independent values of the
harmonic and pure quartic potentials, respectively~\cite{Cretegny}.

In order to smooth out fluctuations, both $\DLE(t)$ and ${\cal L}(t)$
have been averaged over different realizations (typically 20) of 
the equilibrium initial conditions.

\section{\label{sec_harm} The harmonic chain}

In this section we discuss the results obtained for 
the 1D harmonic lattice,
\[
V(x) =  \frac{1}{2} k_2 x^2,
\]
with both free--end  and fixed--end BC. In order to solve the problem, 
we can use ordinary time--dependent perturbation theory, provided the 
damping constant $\g$ is small enough. We note that, since the number 
of the damped particles
is fixed, the perturbative approximation is expected to improve by increasing
the system size $N$.  In the following we shall adopt adimensional units such
that $k_2 = m = a = 1$,  where $a$ is the lattice spacing. 
As a consequence, time is measured in units of $1/\o_0$,  
where $\o_0 = \sqrt{4 k_2 / m}$ is the maximum
frequency of the linear  spectrum.

Let $-\o^2_\a$ and $\eta^\a$ ($\a = 0, 1, \dots N-1$) 
denote the eigenvalues and normalized eigenvectors,
respectively, of the unperturbed hamiltonian problem, 
where
\[
\o_\a^2 = 4\, \sin^2 \left( \frac{q_\a}{2} \right) \quad .
\]
and the wave--number  $q_\a $ is defined in the Appendix (see eqs.
A.6 and A.7)~.

In the spirit of time--dependent perturbation theory we look for solutions
of the form
\begin{equation}
\label{pert_ans}
u_p(t)=\sum_{\a=0}^{N-1} c_\a(t) e^{-i \o_\a t} \eta_p^\a.
\end{equation}
where $\eta_p^\a$ is the $p$-th component of the eigenvector $\eta^\a$.
By substituting  expression~(\ref{pert_ans}) in the equations of motion, 
we can rewrite them in the form of a system of 
differential equations for the $c_\a$'s. We can then 
find approximate solutions by expressing the latter 
as power series in the adimensional perturbation parameter
$\g/\o_0$. The details of the calculation are reported in the Appendix. 
To first order in $\g/\o_0$ we obtain
\begin{equation}
\label{calfa1}
c_\a(t)   = c_\a(0) \exp  \left[-
             \frac{ t}{\tau_\a}\right],
\end{equation}
where
\begin{equation}
\label{Gaaeq_free}
\frac{1}{\tau_\a}  = 
                    \frac{1}{\tau_0}  
                    \, \cos^2 \left(\frac{q_\a}{2} \right),
\end{equation}
and
\begin{equation}
\label{Gaaeq_fix}
\frac{1}{\tau_\a} = 
                     \frac{1}{\tau_0}  
                     \, \sin^2 (q_\a)
\end{equation}
for the free--end and the fixed--end system, respectively,
where $\tau_0 =  N / 2 \gamma$. We also note that,
as a result of the calculation, the damping term introduces a correction
of the normal frequencies. However, it is straightforward to verify that
such a correction is of second order in the perturbation
parameter and, accordingly, it can be neglected.

As we see from equations~(\ref{Gaaeq_free}) and~(\ref{Gaaeq_fix}), the spectra
of the decay rates are substantially different for the free--end and 
fixed--end BC. In the former case, we see that the least damped modes are the
short--wavelength ones ($\a \approx N$), the largest lifetime
being $\tau_{\scs N-1} \approx 2 N^3 / \pi^2 \gamma$, while the most
damped modes are the ones in the vicinity of $\a = 0$, with 
$\tau_0$ being the shortest decay time.  On the contrary, for the
fixed--end chain the most damped modes are the ones around $\a \approx N/2$,
while both the short--wavelength and long--wavelength modes are very little
damped, being  $\tau_{\scs N-1} \approx N (N + 1)^2 / 2 \pi^2
\gamma$. 

Using the definition of the system energy (\ref{energy}) and the solution
(\ref{calfa1}), we can explicitely evaluate the normalized system energy as
\begin{equation}
\label{Toten}
\frac{E(t)}{E(0)} = \frac{\ds \sum_{\a }  c^2_\a(0)\o_\a^2 \, e^{-2t/\tau_\a} }
                          {\ds \sum_{\a}  c^2_\a(0) \o_\a^2}.
\end{equation}
As we are interested in the typical behaviour of the system when it evolves
from equilibrium initial conditions, we replace $ c^2_\a(0) \o_\a^2$ with its 
average value $2E(0)/N$. Thus, recalling equations~(\ref{Gaaeq_free}) 
and~(\ref{Gaaeq_fix}) and approximating for large $N$ the sum over $\a$ with 
an integral ($\tau_\a\to\tau(q)$), we have  
\begin{equation}
\label{Toten_free}
\frac{E(t)}{E(0)} = \frac{1}{\pi}  
                     \int_0^\pi e^{-\frac{\ds 2t}{\ds \tau(q)}} \, dq =
                     e^{-\frac{\ds t}{\ds \tau_{\scs 0}}} 
                     I_{\scs 0} \left( \frac{\ds t}{\ds \tau_{\scs 0}} \right)
\end{equation}
where $I_{\scs 0}$ is the modified zero--order Bessel function. 

Surprisingly enough, although the mode relaxation rates are different 
in the free--end and fixed--end systems, the global result for 
the energy decay turns out to be the same. The asymptotic behaviour of the 
function (\ref{Toten_free}) is given by
\begin{equation}
\frac{E(t)}{E(0)} = \left\{
                        \begin{array}{ll}
                    e^{\ds -t / \tau_{\scs 0}}                               & 
                    \ {\rm for} \ t \ll \tau_{\scs 0},                        \\
                    \frac{\ds 1}{\ds \sqrt{2 \pi (t/\tau_{\scs 0}) }}        &
                    \ {\rm for} \ t \gg \tau_{\scs 0}.                                               
                        \end{array}
                    \right.
\label{asymp}
\end{equation}
Hence, $\tau_{\scs 0}$ sets the time scale of a crossover from the initial
exponential decay, led by the longest lifetime, to a power--law
decay. 

\begin{figure}[ht!]
\centering
\subfigure[]{
\includegraphics[width= 6.3 truecm, clip]{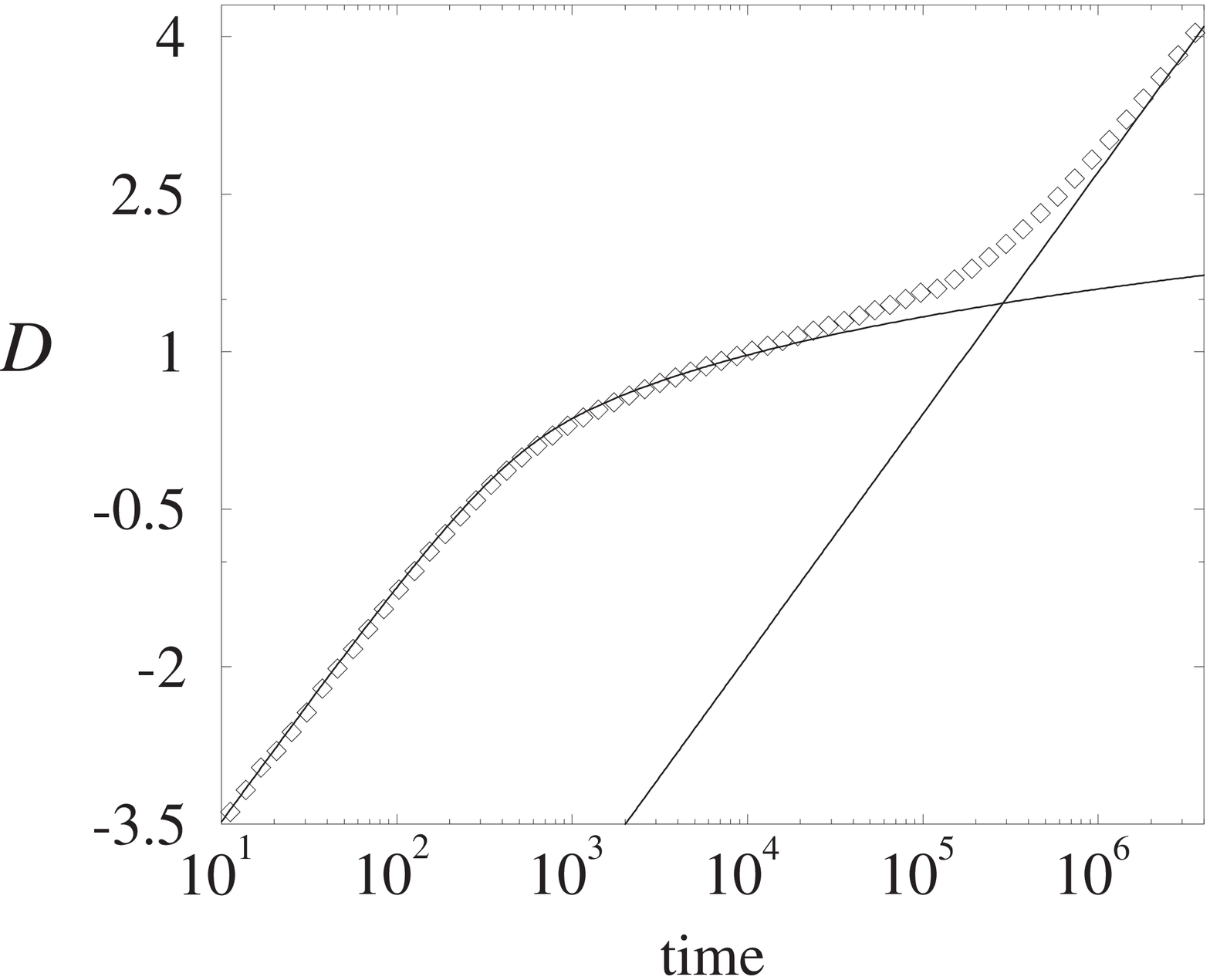}
}
\hspace{.5 mm}
\subfigure[]{
\includegraphics[width= 6 truecm, clip]{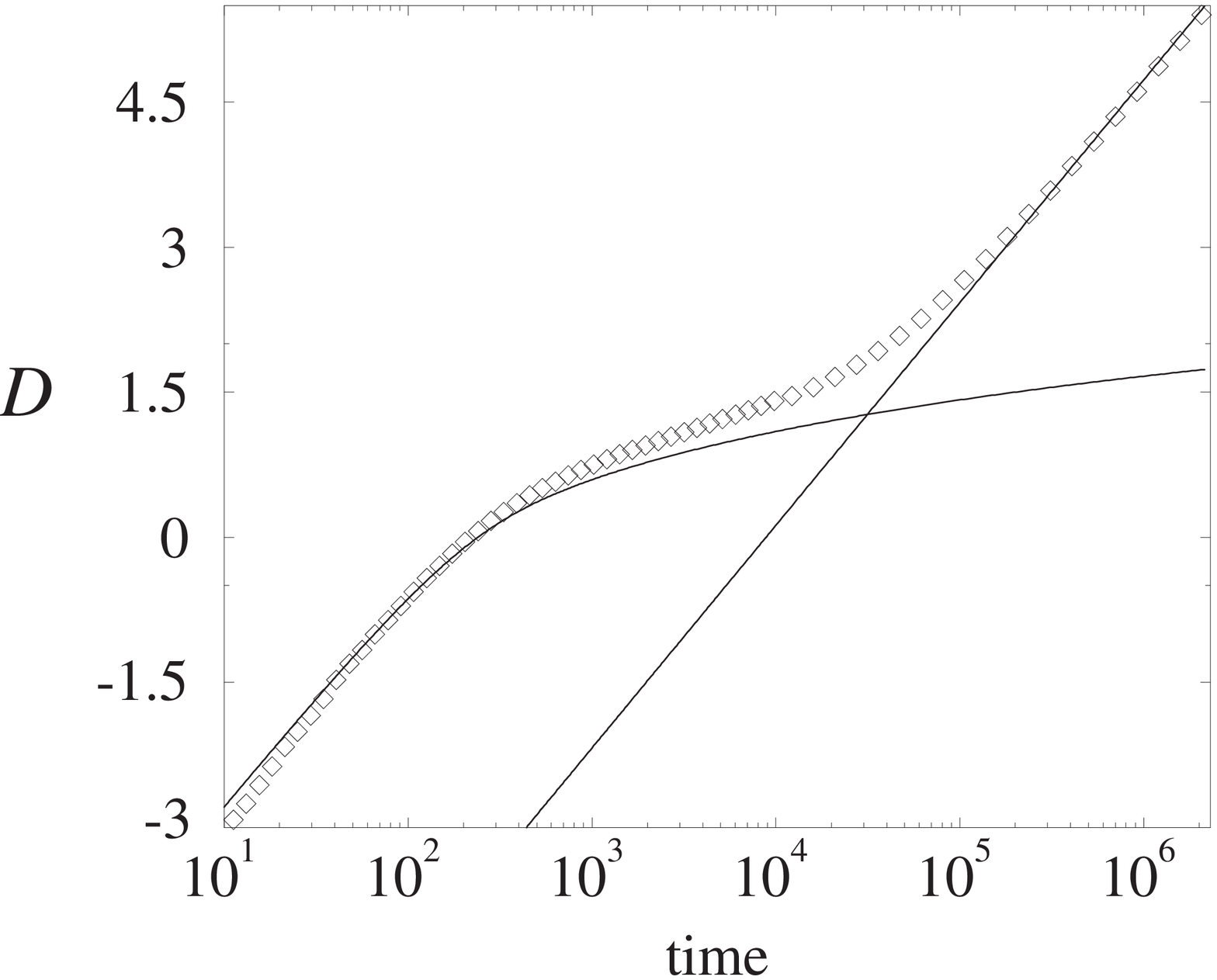}
}
\caption{\em \label{En32} 
Symbols represent $\DLE$ versus time for a harmonic chain with $N = 32$, 
and with free--end {\rm (a)}
and fixed--end {\rm (b)} boundary conditions,
$\gamma=0.05,0.1$, respectively.
Solid lines are plots of 
formula~(\ref{Toten_free}) and of the function $\exp(-2t/\tau_{\scs N-1})$. 
}
\end{figure}

This peculiar behaviour is illustrated in figure~\ref{En32}. In order to check
the validity of the above approximations we compared the  
analytical result (\ref{asymp}) with the outcome of numerical simulations. We
plot in a log--lin scale $\DLE$ vs time for two different 
lattices of size $N=32$ with fixed--end and free--end BC. 
We obtain an excellent agreement at least up to $t\approx \tau_{\scs N-1}$, 
where a further crossover to an exponental decay occurs. 
This is an expected finite--size effect.
Indeed, for times larger than $\tau_{\scs N-1}$  only the
modes around $\a = N-1$ are populated (in the fixed--end chain also
those close to $\a = 0$). Therefore, the long--term behavior of the function
$E(t)/E(0)$ will be exponential, with time constant $\tau_{\scs N-1}/2$. However, we
note that the second crossover is BC--dependent, since the values of
$\tau_{\scs N-1}$, although both proportional to $N^3$, are different in the
two cases. 

To conclude this section, let us remark that the results reported here
apply in their present form also to a more general class of models 
with harmonic on--site potential.

%______________________________  FPU SYSTEM _______________________________

\section{\label{sec_nln} The FPU chain}

In this section we study the FPU model, 
\begin{equation}
\label{eqmot_FPU}
V(x) = \frac{1}{2} x^2 + \frac{1}{4} x^4 ,
\end{equation}
where the coupling constant of the quartic term has been set to one and we
consider the energy as the only independent parameter of the hamiltonian
system.
 
\begin{figure}[ht!]
\centering
\subfigure[]{
\includegraphics[width=7 truecm, clip]{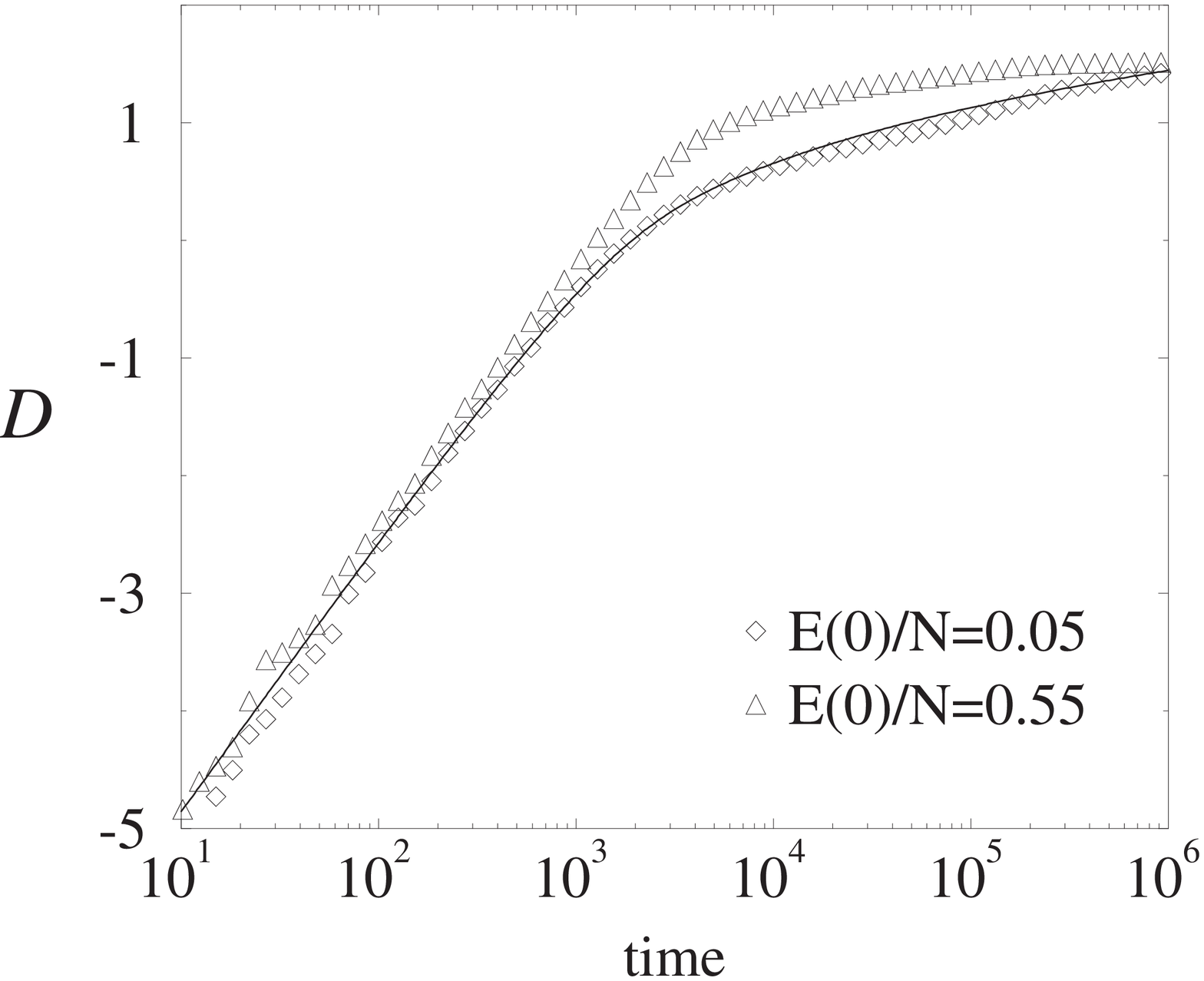}
}
\hspace{.5 mm}
\subfigure[]{
\includegraphics[height=7 truecm, angle=-90, clip]{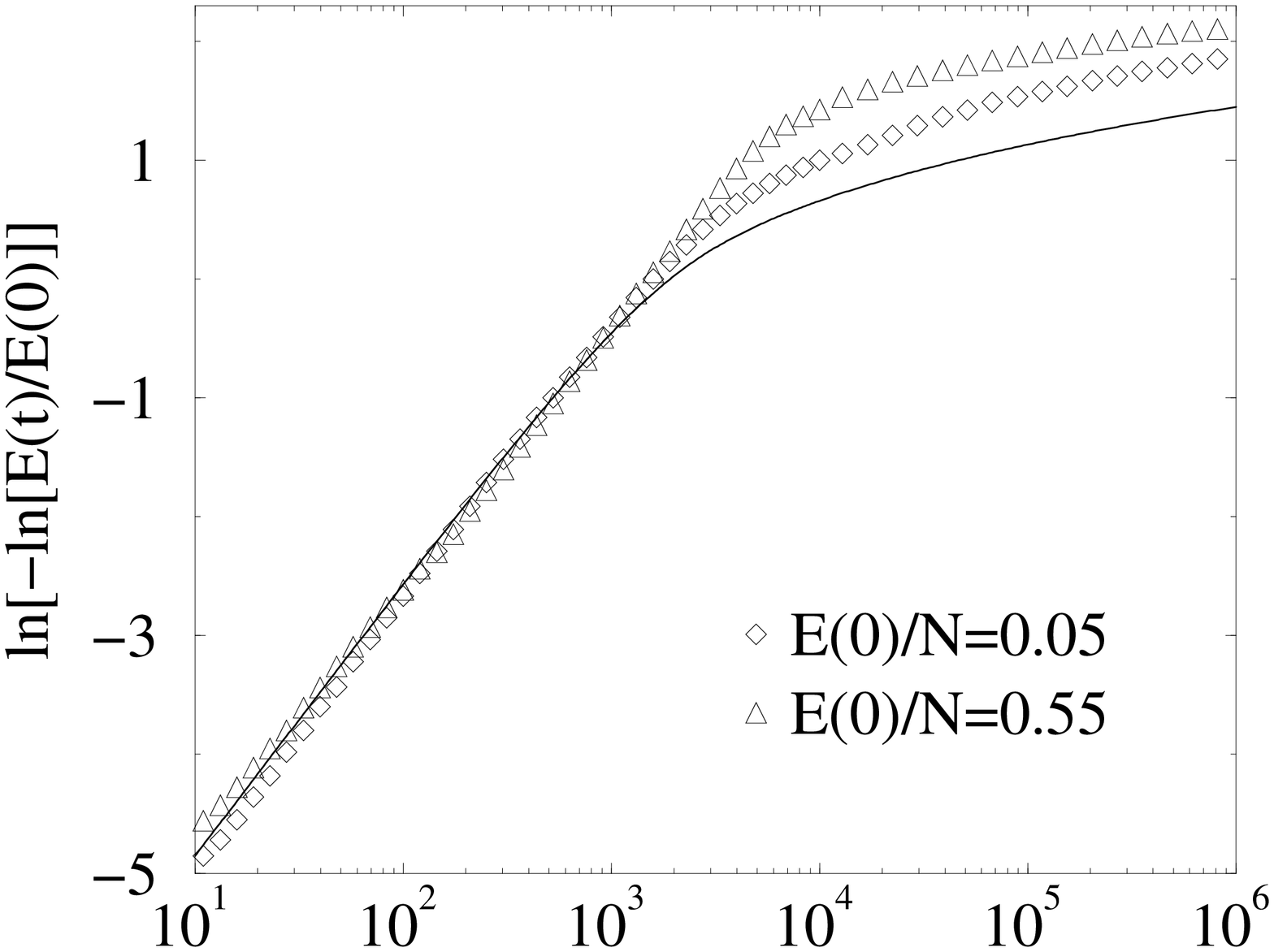}
}
\caption{\em \label{FPU_relax} 
Symbols represent $\DLE$ versus time for an FPU chain with $N = 256$, 
$\gamma=0.05$ and with free--end {\rm (a)}
and fixed--end {\rm (b)} boundary conditions. 
Solid line is the plot of formula~(\ref{Toten_free}). 
}
\end{figure}

In figure~\ref{FPU_relax} we plot $\DLE$ vs time for two chains of size 
$N=256$ with free--end and fixed--end BC. In both cases no 
evidence of stretched exponential decay is found. The curves follow to a very
good extent the characteristic trend of the linear model (see
equation~(\ref{Toten_free})~). This remarkable observation can be interpreted
by the following arguments. First of all, an approximate description in terms
of an effective harmonic model with energy--dependent renormalized frequencies
proved to succesfully account for a number of equilibrium properties of the FPU
chain~\cite{Alabiso,Lepri}. If we assume that the energy extraction rate is
slow enough for the system to evolve in a quasi--static fashion, we may
reasonably expect that a similar scenario as the one described for the harmonic
case emerges. Second, the harmonic approximation becomes increasingly accurate
as time elapses, simply because more and more energy is extracted from the
system by the reservoir.

Actually, the energy--dependent corrections to the linear spectrum
only induce in the energy decay curves small deviations from the
linear case. Indeed, a  closer inspection of figure~\ref{FPU_relax} 
reveals that changing the initial energy density $E(0)/N$
only affects the first crossover region, for both free--end 
and fixed--end BC.

\begin{figure}[ht!]
\centering
\subfigure[]{
\includegraphics[width= 5.5 truecm, angle=-90, clip]{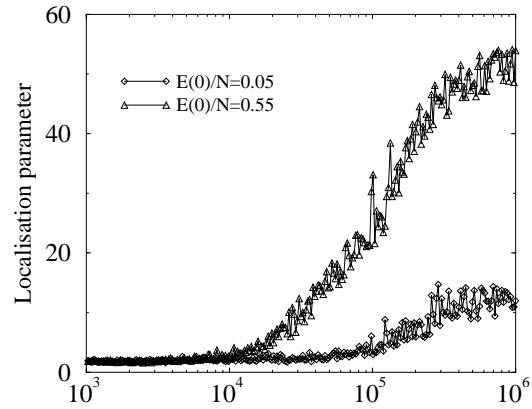}
}
\hspace{.5 mm}
\subfigure[]{
\includegraphics[width= 5.5 truecm, angle=-90, clip]{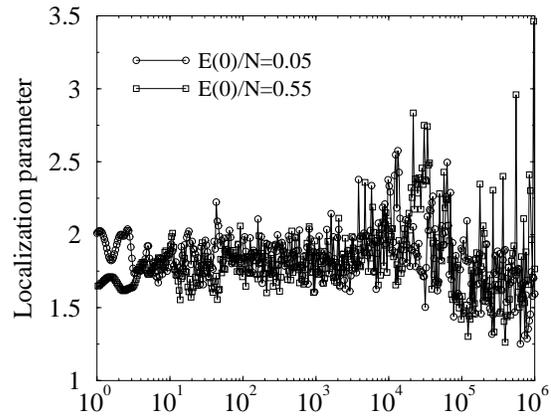}
}
\caption{\em \label{FPU_lp}
The localization parameter ${\cal L}$ vs time for an FPU chain of size $N=256$ 
with free--end {\rm (a)} and fixed--end {\rm (b)} boundary conditions.}
\end{figure}

A dramatic deviation from the linear--like behaviour
emerges however at a later stage, when the localization of energy sets in. 
We observe the spontaneous birth of breathers, analogous
to what reported in ref~\cite{Aubry1}.
This process pins the energy at some sites in the chain, thus forcibly causing 
a further slowing down of the energy decay. Nevertheless, we note that this   
can hardly be detected from the energy decay curves, which are mainly 
determined by the relaxation of the Fourier modes.
On the other hand, localization is clearly revealed by plotting the 
parameter ${\cal L}$ defined by formula~(\ref{local})
(see figure~\ref{FPU_lp}). Remarkably, in the free--end chain
we have observed spontaneous excitation of breathers for all
the performed numerical experiments. On the contrary,
localization turns out to be strongly inhibited for the
fixed--end chain. As a matter of fact, we have observed it
in a very small fraction (say $5 \%$) of the numerical experiments.
Moreover, the {\em degree of localization} (i.e. the asymptotic
value of ${\cal L}$) depends on the initial energy.

\begin{figure}[ht!]
\centering
\subfigure[]{
\includegraphics[height=8 truecm, angle=-90, clip]{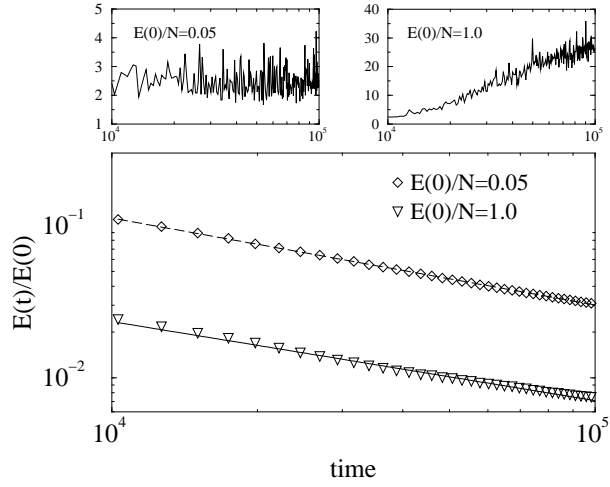}
}
\hspace{.2 mm}
\subfigure[]{
\includegraphics[height=7 truecm, clip]{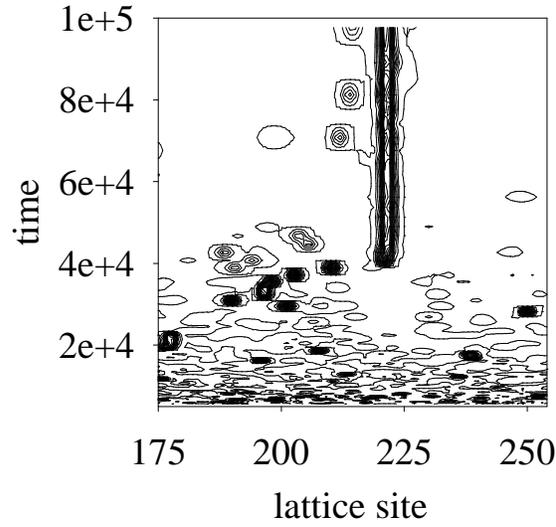}
}
\caption{\em \label{pwl_comp} FPU chain with $N=256$.
{\rm (a)} Lower panel: log-log plot
of the first power--law portion of the energy decay curves.
Best fits (solid lines) give exponents $-0.51$ and $-0.53$ 
for $E(t)/E(0) = 0.05$ and $1.0$, respectively. The upper panels show the 
trend of the localization parameter ${\cal L}$ for both cases in the same 
time domain. 
{\rm (b)}
Space--time contour plot of the symmetrized site energies for 
$E(t)/E(0)=1.0$.
The figure shows a close--up in the region of the chain where 
the breather develops.}
\end{figure}

The pathway to localization is similar to the one described in 
ref~\cite{Cretegny}, whereby one breather is formed from a collection of
short--lived localized vibrations that eventually merge together.  The
situation is shown for a typical simulation in figure~\ref{pwl_comp} (b) in the
form of a space--time contour plot of the symmetrized site energies. 
The localized modes that emerge out of  the
relaxation process are very similar to the ones that originate from the
modulational instability of the zone--boundary modes in the hamiltonian
system~\cite{Cretegny,mod_inst}. This is seen by looking at the displacement
and velocity pattern of the  particles and their spatial spectra. Moreover, the
localized  objects are seen to either stay fixed or move with a definite 
velocity~\cite{mod_inst}. As a consequence they may collide  with the
boundaries, thereby loosing some energy. However, these losses turn out to be
tiny, not only because the interaction with the reservoirs are 
quasi--elastic (at least for $\gamma \ll 1 $) but also because 
the breather velocity is typically smaller  than the sound speed. 

\begin{figure}
\begin{center}
\includegraphics[height=9 truecm, angle=-90, clip]{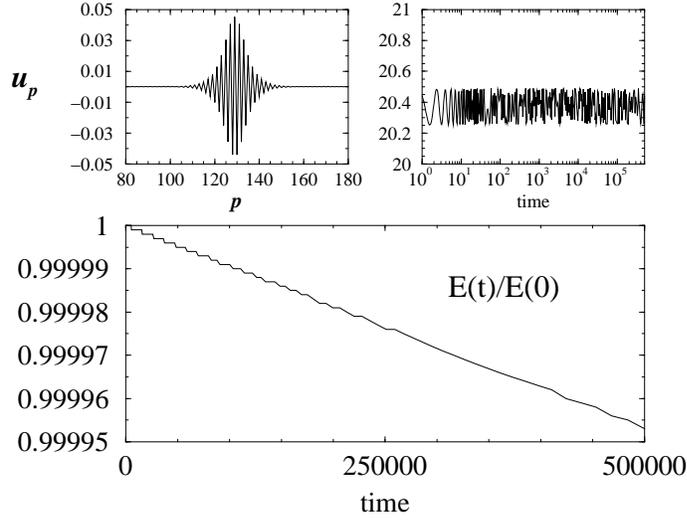}
\caption{\em \label{relax_RWA} Plots of an RWA solution relaxing in an $N=256$
chain. Lower panel: normalized energy vs time. 
Upper left panel: snapshot of 
the atomic displacements at $t=5 \cdot 10^6$. 
Upper right panel: the localization parameter ${\cal L}$ vs time.}
\end{center}
\end{figure}

To further confirm that the damped system supports breathers similar  to
the ones known from the hamiltonian case, we have performed the following
numerical experiment. A pattern corresponding to a  
zero--velocity DB solution in a lattice of size $N = 256$ has been
computed numerically within the framework of the rotating wave
approximation  (RWA)~\cite{RWA}. This solution has then been used as the
initial condition for the evolution equations~(\ref{eq_mot}). The results
are summarized in  figure~\ref{relax_RWA}, where we plot the energy decay
curve alongside  with a snapshot of the lattice displacements at the end
of the simulation  and the localization parameter. The dissipation
acts effectively in eliminating  the vibrational junk radiating from the
approximate solution and stabilizing it. The extremely slow decay of
the breather energy (see lower panel in figure~\ref{relax_RWA}) is
presumably due to anelastic scattering with small amplitude
plane waves~\cite{Johansson}. 

We can understand what difference the BC make for localization by
recalling what we have learnt from the harmonic chain together with
the known effect of modulational instability of the 
zone--boundary modes~\cite{mod_inst,Peyrard}. In fact, we see
that the main difference between the two cases is in the  relaxation rates
of the Fourier modes (see equation~(\ref{Gaaeq_free}) 
and~(\ref{Gaaeq_fix})). This implies that in the free--end system the
long--wavelength modes quickly disappear from the lattice, 
leaving there just the long--lived zone--boundary ones. 
This explains why modulational instability is so 
effective for the formation of breathers.  On the contrary, the 
``longevity'' of the long--wavelength modes in the fixed--end chain spoils
such instability process, unless a very big energy fluctuation overcomes
the other effect  providing an alternative localization pathway.

\begin{figure}
\begin{center}
\includegraphics[height=8.5 truecm, clip]{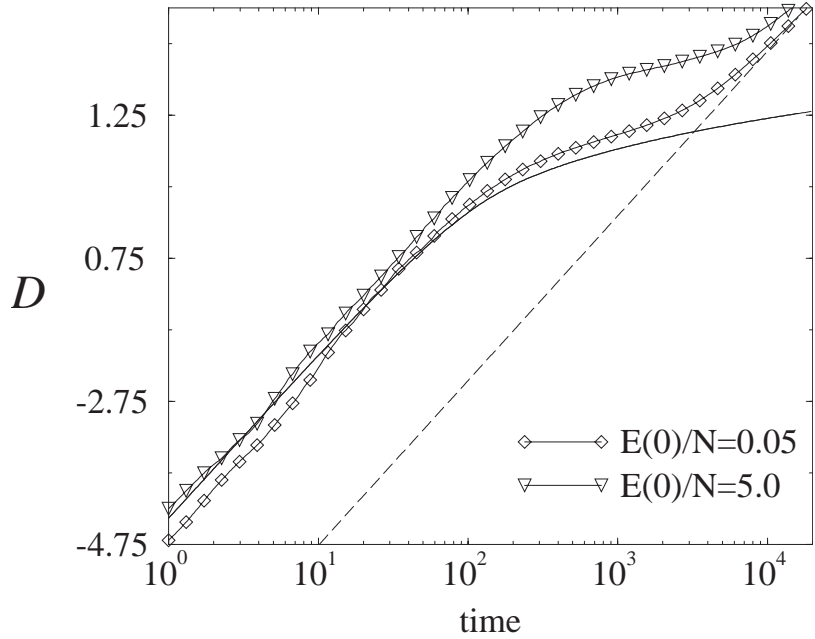}
\caption{\em \label{fix_long} 
Symbol represent $\DLE$ versus time for an FPU chain with
$N=16$ and fixed--end boundary conditions. 
Solid and dashed lines are plots of formula~(\ref{Toten_free}) and
of the law $\exp(-2t/\tau_{N-1})$, 
with $\tau_{N-1}=N(N+1)^2/2 \pi^2 \g$, respectively.}
\end{center}
\end{figure}

As already pointed out, only the long--term portion
of the energy relaxation curves is affected by
localization. Indeed, in the absence
of breathers, the energy slowly drifts back to an exponential law.
This is clearly seen in figure~\ref{fix_long}, where we
plot $\DLE$ versus time for a fixed-end FPU chain.
In order to better illustrate this point, we show in figure~\ref{pwl_comp} (a)
two fits to the first power--law portion of the energy decay curves
for two values of the initial energy. In the lower energy simulation
localization has not yet occurred. On the contrary, at the higher energy, 
the same portion of the curve corresponds
to arousal of localization (see also figure~\ref{pwl_comp} (b)). 
This is reflected in the trend of ${\cal L}$ for the two cases.
We see that the trends of $\DLE(t)$ are clearly indistinguishable 
in the two cases, making emergence of localization difficult 
to spot from the energy decay curves.

\section{\label{conclu} Conclusions}

We have analysed the energy relaxation of chains of atoms with linear and
FPU interaction potential by applying a damping term at the chain boundaries.
Both free--end and fixed--end conditions have been considered.  
We have shown that also the simple linear system is characterized 
by nontrivial relaxation features, that are determined by
the interplay of different relaxation time scales. These are
characteristic for each Fourier mode,
which relaxes exponentially in time with a  decay rate
depending on its wavenumber.
In particular, we have derived a two--crossover picture of the decay
process.  More precisely, for $t > \tau_0 = N/2 \g$ the initial exponential 
behaviour slows down to an inverse square--root law. 
At a later stage ($t > \tau_{\scs N-1} \propto N^3 / \g$)
the system comes back to an exponential decay with a relaxation rate
twice the one of the slowest Fourier mode.
Furthermore, the first crossover is independent of the BC, while the 
second one depends on them. 

We have analysed the nonlinear system under the same conditions and found that
the behaviour is almost concident with the linear case, provided the energy 
extraction rate is not too fast. Noticeably, our data are not compatible 
with a stretched--exponential relaxation. As a side but important result, 
we clarify why the fixed--end FPU chain does not 
display spontaneous localization as a product
of the energy extraction  process. This is because
the long--wavelength modes of the fixed--end chain at equilibrium
are characterized by high amplitudes and vanishing 
relaxation rates, that are proportional to $\g/N^3$~. 
As a result, the modulational instability of the short--wavelength modes
is not effective in producing localized vibrations and 
the energy decay behaviour of the fixed--end FPU chain is practically
equivalent to the linear case.

The dependence of the characteristic time scales on the system size
stems from the choice of the damping matrix $\G$. 
More generally, if one fixes the fraction $f$ of the damped particles,
the first crossover time $\tau_0$ becomes independent of $N$ 
(and inversely proportional to $f$), while
the second one $\tau_{\scs N-1}$ is still expected to scale as $N^3$.

In summary, we have brought forward the intrinsic complexity
of spontaneous breather excitation in spatially
extended nonlinear systems.
In particular, we have illustrated the importance of the system size
and of the boundary conditions for such studies.

We want also to observe that the above conclusions hold
for 1D nonlinear chains, where relaxation produces 
localization of the energy for arbitrarily small initial energy. 
The situation may be significantly different in 2D due to the 
presence of an energy threshold for breather solutions~\cite{2D_tresh}.
Preliminary numerical studies of the 2D 
FPU model seem to confirm such expectations.
This point certainly deserves deeper investigations, that might
provide further insight on the relevance of breathers
as spontaneous excitations emerging from a relaxation dynamics in
nonlinear lattices.

\ack

F.P. warmly aknowledges various helpful discussions with L. Cianchi and 
P.Moretti, and is also indebted to the research group {\it Dynamics of 
Complex Systems} in Florence for the kind hospitality. This work is 
part of the INFM project {\it Equilibrium and nonequilibrium dynamics 
in condensed matter}.      

\appendix
\section[A]{}

In this Appendix we report the details of the calculation of the damping rates
for the harmonic lattice. In matrix notation the equations of motion read
\begin{equation}
\label{eqmot}
\ddot{u} = K u - \Gamma \dot{u}
\end{equation}
where $u$ is the column vector $(u_0, u_1, \dots u_{N-1})$, while $K$
is  the matrix
\begin{equation}
\label{Kappa}
K_{p p^{\pr}} = \frac{k_2}{m} \left[ \d_{p^{\pr},p+1} + \d_{p^{\pr},p-1}
                                     - 2 \d_{p^{\pr},p} \right] 
\end{equation}
The top and bottom diagonal elements of matrix $K$ are defined as
$K_{00} = K_{(N-1),(N-1)} = - k_2/m$ for free--end BC, while definition
~(\ref{Kappa}) applies to fixed--end BC. The eigenvectors of the 
hamiltonian problem satisfy the orthonormality conditions
\begin{eqnarray}
\label{ortho}
\eqalign{
(\eta^{\a}, \eta^{\a^{\pr}} ) = \sum_{p=0}^{N-1} \eta^\a_p 
                                 \eta^{\a^{\pr}}_p =
                                 \d_{\a,\a^{\pr}} \\
(\eta^{\a}, K \eta^{\a^{\pr}} ) = \sum_{p,p^{\pr}=0}^{N-1} 
                              \eta^\a_p K_{p p^{\pr}} \eta^{\a^{\pr}}_{p^{\pr}}
                               = -\o_\a^2 \d_{\a,\a^{\pr}}, 
}
\end{eqnarray}
By substituting  expression~(\ref{pert_ans}) in equation~(\ref{eqmot}), 
we get
\begin{eqnarray*}
\fl
\sum_{\a=0}^{N-1} 
      [\ddot{c}_\a(t) -\o_\a^2 c_\a(t) -2 i \o_\a \dot{c}_\a(t)] 
      e^{-i \o_\a t} \eta^\a_p -
\sum_{\a=0}^{N-1} c_\a(t) e^{-i \o_\a t} 
\sum_{p^{\pr}=0}^{N-1} 
      K_{p p^{\pr}} \eta^\a_{p^{\pr}} = \\ 
 - \sum_{\a=0}^{N-1} 
      [\dot{c}_\a(t) - i \o_\a c_\a(t)] e^{-i \o_\a t}
\sum_{p^{\pr}=0}^{N-1}
      \Gamma_{p p^{\pr}} \eta^\a_{p^{\pr}}.
\end{eqnarray*}
By taking the scalar product of the latter equation with $\eta^{\a^{\pr}}$ and
recalling the orthonormality conditions contained in (\ref{ortho}), 
we are led to the system of equations
\begin{equation}
\label{eqbeta1}
\fl
\ddot{c}_\a(t) - 2 i \o_\a \dot{c}_\a(t) = 
                     - \sum_{\a^{\pr}=0}^{N-1} 
                       [\dot{c}_{\a^{\pr}}(t) - i \o_{\a^{\pr}} 
                        c_{\a^{\pr}}(t)] \, \G^{\a \a^{\pr}}
                       e^{-i (\o_{\a^{\pr}} -\o_\a)t},
\end{equation}
where
\begin{equation}
\label{Gamma_aa}
\Gamma^{\a \a^{\pr}} = (\eta^{\a}, \Gamma \, \eta^{\a^{\pr}} ) = 
                       \sum_{p,p^{\pr}=0}^{N-1} 
                       \eta^\a_p \Gamma_{p p^{\pr}} \eta^{\a^{\pr}}_{p^{\pr}}
\end{equation}
We can obtain approximate solutions of  system~(\ref{eqbeta1}) by means
of a perturbative series of the form
\[
c_\a(0) = c_\a^{[0]}(t) + c_\a^{[1]}(t) + c_\a^{[2]}(t) + \dots,
\]
where $c_\a^{[1]}(t)$, $c_\a^{[2]}(t)$, $\dots$  are first order, 
second order amplitudes, and so on, in the perturbation parameter $\g$. 
We can then use the usual iterative procedure. If we assume that at $t=0$ 
only the mode $\a=\bar{\a}$ is populated, we can approximate 
$c_\a(t) = \d_{\a,{\bar{\a}}}$ (independent of $t$) in the right--hand 
side of equation~(\ref{eqbeta1}) and then integrate to get $c_\a^{[1]}(t)$.
The procedure can be repeated for obtaining $c_\a^{[2]}(t)$, etc.
For the first order approximation  we obtain the following equation
\[
\ddot{c}_\a(t) - [ 2 i \o_\a -  \G^{\a \a} ] \dot{c}_\a(t)
                         - i \o_\a  \G^{\a \a} c_\a(t) = 0,
\]
which is readily integrated, yielding expression~(\ref{calfa1}).

To calculate the matrix elements $\G^{\a \a}$, we need
the explicit expression of the eigenvectors of the unperturbed  
problem, which are
\begin{equation}
\label{evc_free}
\fl
\eta^\a_p = \sqrt{\frac{2}{N}} \cos 
                                        \left[ 
                                        q_\a \left( p +\frac{1}{2} \right)
                                        \right], \ \ \ \ q_\a = \frac{\a \pi}{N},
                                        \ \ \ \ \a = 0,1 \dots N-1
\end{equation}
for the free--end chain, and
\begin{equation}
\label{evc_fix}
\fl
\eta^\a_p = \sqrt{\frac{2}{N}} \sin 
                          \left[ 
                          q_\a (p + 1)
                          \right], \ \ \ \ q_\a = \frac{(\a +1) \pi}{N+1},
                          \ \ \ \ \a = 0,1 \dots N-1
\end{equation}
for the fixed--end one.
By inserting these expressions in equation~(\ref{Gamma_aa}), one can
easily obtain equation~(\ref{Gaaeq_free}) and equation~(\ref{Gaaeq_fix}).

\Figures

\end{document}